# GenDMR: A dynamic multimodal role-swapping network for identifying risk gene phenotypes

Lina Qin[1], Cheng Zhu[1], Chuqi Zhou[1], Yukun Huang[2], Jiayi Zhu[3], Ping Liang[1], Jinju Wang[4], Yixing Huang[5], Cheng Luo[6], Dezhong Yao[6], Ying Tan[1*]

[1]The Key Laboratory for Computer Systems of State Ethnic Affairs Commission, Southwest Minzu University, Chengdu, 610225, China

[2]School of Education and Psychology, Southwest Minzu University, Chengdu, 610225, China

[3]State Key Laboratory of Cognitive Neuroscience and Learning, Beijing Normal University, Beijing, 100875, China

[4]School of mathematics, Southwest Minzu University, Chengdu, 610225, China

[5]Institute of Medical Technology, Peking University Health Science Center, Beijing, 100191, China

[6]The Clinical Hospital of Chengdu Brain Science Institute, MOE Key Lab for Neuroinformation, School of Life Science and Technology, University of Electronic Science and Technology of China, Chengdu, 611731, China

## Abstract.

Recent studies have shown that integrating multimodal data fusion techniques for imaging and genetic features is beneficial for the etiological analysis and predictive diagnosis of Alzheimer's disease (AD). However, there are several critical flaws in current deep learning methods. Firstly, there has been insufficient discussion and exploration regarding the selection and encoding of genetic information. Secondly, due to the significantly superior classification value of AD imaging features compared to genetic features, many studies in multimodal fusion emphasize the strengths of imaging features, actively mitigating the influence of weaker features, thereby diminishing the learning of the unique value of genetic features. To address this issue, this study proposes the dynamic multimodal role-swapping network (GenDMR). In GenDMR, we develop a novel approach to encode the spatial organization of single nucleotide polymorphisms (SNPs), enhancing the representation of their genomic context. Additionally, to adaptively quantify the disease risk of SNPs and brain region, we propose a multi-instance attention module to enhance model interpretability. Furthermore, we introduce a dominant modality selection module and a contrastive self-distillation module, combining them to achieve a dynamic teacher-student role exchange mechanism based on dominant and auxiliary modalities for bidirectional co-updating of different modal data. Finally, GenDMR achieves state-of-the-art performance on the ADNI public dataset and visualizes attention to different SNPs, focusing on confirming 12 potential high-risk genes related to AD, including the most classic APOE and recently highlighted significant risk genes. This demonstrates GenDMR's interpretable analytical capability in exploring AD genetic features,

---

[*] Corresponding author.

providing new insights and perspectives for the development of multimodal data fusion techniques.

# 1. Introduction

Alzheimer's disease (AD) is a progressive neurodegenerative disorder primarily characterized by memory impairment and cognitive decline. Currently, there is no cure for AD, and accurate diagnosis plays a crucial role in improving patient survival rates [1]. Structural magnetic resonance imaging (sMRI) is a non-invasive imaging technique with high spatial resolution, capable of precisely revealing structural changes in brain tissue. It effectively captures brain atrophy commonly observed in AD patients [2]. However, approximately 80% of AD cases are believed to have a genetic basis [3]. Relying solely on neuroimaging often falls short in uncovering the genetic mechanisms underlying AD, leading to insufficient biological interpretability of the disease. Consequently, integrating genetic and imaging data—an approach known as imaging genetics—has emerged as a promising research direction.

Nevertheless, not all brain changes are driven by genetic factors, making it unreasonable to indiscriminately apply genetic features in AD diagnosis and prediction. Against this backdrop, deep learning models based on risk-related genetic variants, specifically single nucleotide polymorphisms (SNPs), have garnered increasing attention. Relevant studies advocate for first identifying high-risk SNPs, followed by their integration into deep learning frameworks for multimodal fusion. For example, Wang et al. proposed a cross-modal diagnostic framework combining SNPs and imaging data using graph diffusion and hypergraph regularization [4]; Ko et al. jointly optimized SNP embeddings and phenotypic data for both disease classification and cognitive score prediction [5]. However, these methods still overlook two critical issues:

First, the biological validity of current SNP encoding strategies remains inadequate. Most existing approaches represent each SNP using discrete numerical values (e.g., {0, 1, 2}), treating them as independent entities without considering their relative positions. This simplified representation contradicts the well-established concept of linkage disequilibrium (LD) in genetics [6], which posits that SNPs located in close physical proximity on a chromosome often exhibit non-random associations due to shared inheritance or functional coordination. Independent encoding may sever these spatial correlations, potentially leading to the loss of important functional information. This limitation is prevalent across many recent deep learning studies, suggesting that improving SNP representation may be key to enhancing model performance. Therefore, in our subsequent work, we incorporated chromosomal spatial information to effectively model these spatial dependencies.

Second, current SNP interpretability analyses lack adaptive mechanisms. Most existing approaches rely on manually designed features to identify risk-associated SNPs, lacking an objective and quantifiable analytical framework. As a result, the reliability of such analyses is often limited. Although some studies have attempted to integrate SNP data with deep learning models, the inherent "black-box" nature of these models hinders biological interpretability [7]. Consequently, few models are able to demonstrate the contribution of individual SNPs through visualizations of adaptive parameters. To address this gap, we propose an adaptive SNP scoring strategy based on a multi-instance learning (MIL) framework—a novel deep learning paradigm [8]. This approach enables the visualization of each SNP's contribution to AD risk.

On the other hand, each modality possesses unique value and the contribution of relatively weaker modalities should not be overlooked. Currently, classification models based on sMRI images typically achieve accuracy rates exceeding 80%, whereas SNP-based models often fall below 70% [9]. This indicates that sMRI features contribute more significantly to classification performance than SNP features. In response to such modality imbalance, recent studies have employed knowledge distillation techniques, generally treating the multimodal model as the teacher and the dominant unimodal feature (e.g., imaging) as the student [10, 11]. However, relatively weaker modalities such as SNPs still contain distinctive information. We argue that SNPs can also serve as teachers to compensate for the limitations of sMRI, and in doing so, can help uncover the differential importance of individual SNPs [12]. Based on this insight, we propose a novel **dynamic teacher-student role exchange mechanism** to facilitate mutual learning between modalities.

To address the aforementioned challenges, we propose a novel **Dynamic Multimodal Role-swapping Network for Genetics and Images (GenDMR)**. First, recognizing the limitations in the biological validity of conventional SNP encoding methods, we enhance SNP feature representation by incorporating spatial information and employing a multi-instance partitioning strategy at the SNP site level. This approach captures both the heterogeneity and interdependence among SNPs. Building on this, we design a **multi-instance attention module** that adaptively quantifies the contribution of individual SNP sites to Alzheimer's disease (AD), thereby enhancing model interpretability.

Secondly, we introduce a **Dynamic Teacher-Student Role Exchange Mechanism**, which enables bidirectional co-updating between modalities by dynamically adjusting the flow of information during training. Specifically, we propose a **dominant modality selection module** that quantifies feature uncertainty (via entropy) and the discriminative capacity of each modality (via U-shaped distribution scoring) to adaptively identify the most informative modality at each epoch. Finally, inspired by the concept of self-distillation [13], we develop a **contrastive self-distillation module**, in which the selected dominant modality acts as the teacher and the auxiliary modality as the student, enabling dynamic fusion of multimodal information.

**The main contributions of this study are as follows:**
1. **Improved SNP Encoding in Deep Learning Models**: We introduce spatial

indicators to effectively model the spatial relationships among SNPs. Moreover, we adopt a gene-level multi-instance partitioning strategy that captures synergistic effects among SNPs. Combined with the proposed multi-instance attention module, the model achieves interpretable SNP representations by adaptively scoring their disease relevance.

2. **Dynamic Teacher-Student Role Exchange Mechanism**: Acknowledging the differing diagnostic values across modalities in AD, we design a dynamic role-swapping mechanism composed of a dominant modality selection module and a contrastive self-distillation module. This enables **bidirectional co-updating** of modality roles and offers new insights into multimodal data fusion.

3. **Interpretability and Risk SNP Identification**: GenDMR visualizes the risk levels of individual SNPs in relation to AD and reliably identifies 12 potential high-risk AD-related SNPs. These include the well-established **APOE** gene and several emerging risk genes that have gained attention in recent years. This demonstrates GenDMR's capability to identify potential AD risk genes and serves as a useful tool for aiding pathogenic gene assessment.

## 2. Related Work

### 2.1. SNP encoding method

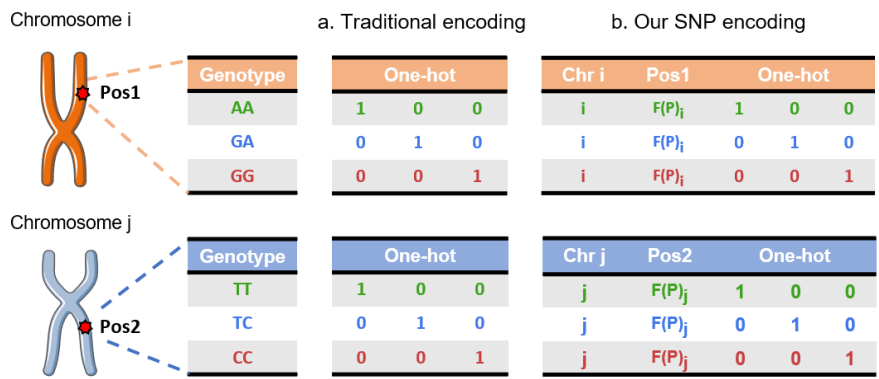

**Fig. 1**. Comparison of SNP encoding methods. The one-hot encoding representation for different genotypes may be same, leading to potential confusion in traditional encoding. Icon credit: Chromosome i and chromosome j icons provided by Servier, from https://smart.servier.com/.

There are various ways to encode SNPs, and different encoding strategies can significantly affect the performance of deep learning models. Therefore, selecting an appropriate encoding method is crucial for downstream analysis [14, 15, 16]. Among current mainstream approaches, SNPs are typically represented either as numerical values or one-hot vectors [4, 17, 18]. As illustrated in Fig. 1, these methods often treat each SNP as an independent entity, without incorporating spatial or contextual information. As a result, the model may fail to distinguish between different genotypes—for instance, treating genotypes AA, GG, TT, and CC as essentially equivalent. To address this limitation, we consider utilizing both the chromosome index and the physical location of each SNP on the chromosome to capture the relationships among different SNPs. Specifically, we adopt the following steps for encoding:

1. Define Chromosome Numbering as $Chr_i$. Where $i \in \{1,2,...,M\}$ represents different chromosomes, $M$ denotes the total number of chromosomes.

2. Let $P \in \mathbb{R}^+$ denote the original position of the SNP. To account for scale differences across chromosomes, $P$ is normalized to the range [0, 1] using the following formula:

$$F(P) = \frac{P - P_{min}}{P_{max} - P_{min}} \quad (1)$$

where $P_{min}$ and $P_{max}$ are the minimum and maximum SNP positions on the chromosome, respectively.

3. The genotype state $S_k$ takes three values (no minor allele, one minor allele, two minor alleles), corresponding to the set $G = \{v1, v2, v3\}$. It is represented as a one-hot vector:

$$g_k = [\delta(S_k = v_1), \delta(S_k = v_2), \delta(S_k = v_3)] \in \mathbb{R}^3 \quad (2)$$

where $\delta(\cdot)$ denotes the Kronecker function, which equals 1 when the condition is satisfied, and 0 otherwise.

In summary, the final SNP encoding vector $SNP_{encoding}$ consists of three components: chromosome number, position, and genotype. It is represented as:

$$SNP_{encoding} = Chr_i \oplus F(P) \oplus g_k \quad (3)$$

where $\oplus$ denotes the direct sum decomposition in Banach space, i.e., the concatenation of features from different feature spaces.

## 2.2. Multi instance learning

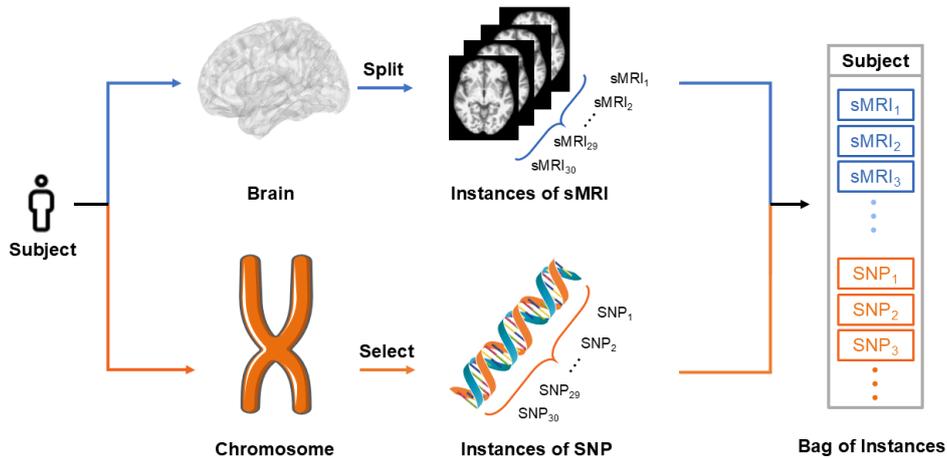

**Fig. 2**. Multi-instance partitioning. One subject corresponds to one bag, and the features of the subject form multiple instances. Icon credit: DNA icon by Servier, from https://smart.servier.com/.

Recent studies have demonstrated that the "bag-instance" structure of Multiple Instance Learning (MIL)—where a single sample is treated as a bag composed of multiple instances derived from its features—can effectively handle label uncertainty in medical imaging tasks [19]. In brain disease diagnosis, label ambiguity is particularly common. Specifically, although each sample typically has a single global label (e.g., benign or malignant), the exact pathological regions or contributing factors are often unclear, and there are no definitive labels for brain features. Thanks to its ability to operate at the bag level without requiring instance-level annotations, MIL is well-suited to this type of uncertainty. It has shown distinct advantages in multimodal disease classification task [20, 21].

In this method, we are given a set of labeled bags, where each bag contains multiple unlabeled instances. A bag is assigned a positive label if at least one instance in the bag is positive; otherwise, the bag is labeled negative if all instances are negative [22, 23]. This logic can be formulated as follows:

$$y_i = \begin{cases} 1 & if \ \sum_j y_i^j \geq 1 \\ 0 & otherwise \end{cases} \quad (4)$$

where $y_i$ denotes the label of the bag (1 for positive, 0 for negative), $y_i^j$ represents the latent label of the $j$ instance in the bag, and the $\sum_j y_i^j \geq 1$ indicates that the bag is positive if at least one instance is positive.

As illustrated in Fig. 2, we apply multi-instance partitioning not only at the image level but also to SNP data. The pathological changes associated with Alzheimer's disease (AD) are typically global, involving varying degrees of brain atrophy or functional abnormalities across regions [24]. A single sMRI slice captures only a limited view, whereas treating a series of slices as instances allows the model to learn more comprehensive patterns of brain alterations. Similarly, complex diseases like AD and depression are driven by the subtle, collective effects of multiple genetic loci. The effect of an individual SNP may be minimal, but the combined influence of multiple SNPs can significantly impact disease risk [25, 26]. By treating SNP loci as instances, the model is better positioned to capture these synergistic genetic effects. Building on this theoretical foundation, we treat each subject as a bag, with their corresponding series of sMRI slices and SNP loci serving as instances. This design aims to more effectively associate brain imaging and SNP data for robust disease modeling.

## 2.3. Knowledge distillation

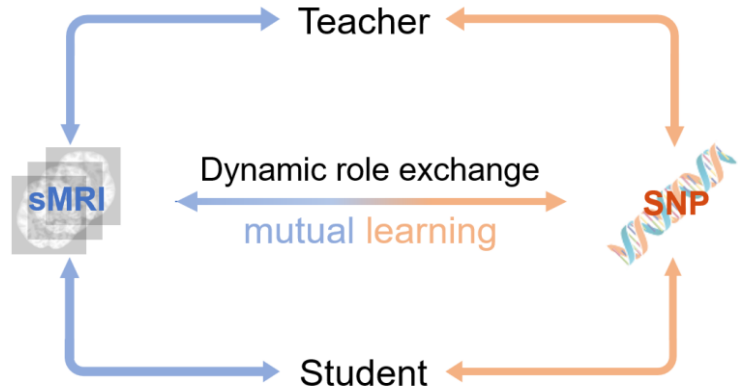

**Fig. 3**. Dynamic Teacher-Student Role Exchange Mechanism. The dominant modality acts as the teacher, and the auxiliary modality as the student. The roles of the modalities dynamically change during training.

To better integrate SNP and sMRI data, we adopt a self-distillation approach. Self-distillation, a variant of knowledge distillation, differs from the traditional teacher-student framework in that the teacher and student models often originate from different layers, branches, or training stages of the same model. Through the transfer of "dark knowledge" embedded in the teacher's representations, the student model is progressively enhanced, ultimately leading to improved overall performance [27]. However, in most existing studies, the roles of teacher and student models are predetermined and remain fixed throughout training—the teacher continuously guides the student, while the student has no influence on the teacher's parameter updates. For example, Xing et al. proposed a two-stage multimodal distillation framework using pathology slides and genomic data. In their approach, a teacher model trained on multimodal inputs provides supervision to a student model that is trained solely on unimodal pathology slide data [11]. Similarly, Wang et al. trained a teacher model using both ocular images and critical clinical features related to cortical opacity, which then guided a student model that only receives ocular image input [10].

In contrast to the aforementioned approaches, we focus on the **bidirectional co-updating** relationship between teacher and student, where no modality is statically assigned as either the teacher or the student throughout training. It is important to emphasize that although sMRI exhibits significant advantages in AD diagnosis, SNP data also contains valuable and unique information. Therefore, allowing SNPs to act as the teacher at certain stages to guide sMRI learning is both reasonable and beneficial.

As shown in Fig.3, we propose a **dominant modality selection module** that dynamically adjusts teacher-student roles during training. The model selects the dominant modality by evaluating the discriminative capacity of each modality and inter-modality uncertainty, thereby enabling adaptive cross-modal contribution weighting. Meanwhile, combined with the contrast self-distillation module improved based on the self-distillation framework, the distance between teachers and students is regulated through the contrast between modes. The core idea is to establish a dynamic teacher-student relationship between different modalities and leverage contrastive learning to tightly map similar instances together while pushing dissimilar instances apart. This mechanism enhances the effectiveness of multimodal fusion and improves the discriminative power of the model.

## 3. Method

In this section, we present GenDMR to learn the features of sMRI and SNP. Our network framework mainly consists of four parts: Feature Generation (FG), Multi-Instance Attention Module (MAM), Dominant Modality Selection Module (DMSM), Contrastive Self-distillation Module (CSM), and Classification Module (CM), as shown in Fig.4.

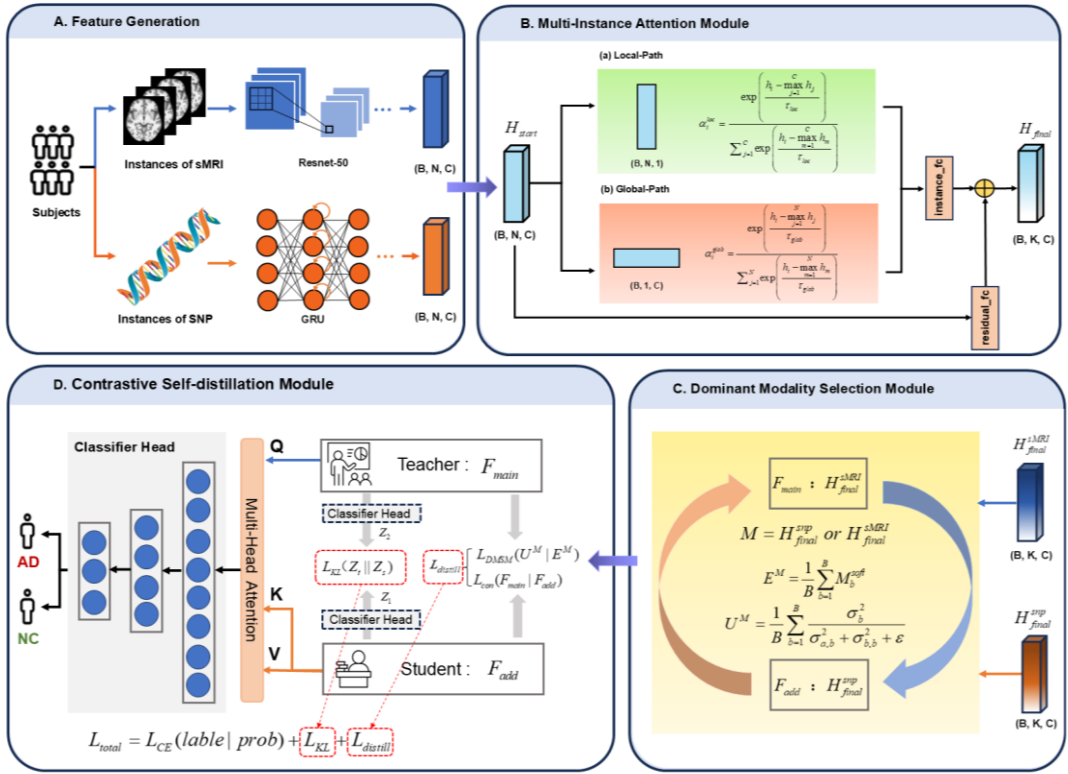

**Fig. 4. Framework of our method.** (A) SNP data is processed using a Gate Recurrent Unit (GRU) [28], and sMRI features are extracted using ResNet-50 [29]. The features from both modalities serve as inputs to the subsequent model. (B) A Multi-Instance Attention Module (MAM) is used to calculate the weights of SNP instances to capture key SNPs. (C) A Dominant Modality Selection Module (DMSM) determines the dominant and auxiliary modalities between the two modalities. (D) A Contrastive Self-distillation Module (CSM) and Classification Head (CH) are used to enhance the matching of features from different modalities in a unified latent space through contrastive learning, improving the model's discriminative power.

## 3.1. Feature Generation

MRI and SNP carry structural imaging information and genetic signals, respectively. Due to significant differences in their representational spaces and sequential structures, direct fusion may lead to modality inconsistency and learning bias. To address this issue and obtain modality representations suitable for downstream fusion and classification, we design a **Feature Generation module (FG)**.

For the sMRI modality, the input consists of continuous brain structural slices, denoted as $X_{sMRI} \in \mathbb{R}^{N \times C \times H \times W}$, where $N$ represents the number of slices (instances), $C$ is the number of channels, and $H$, $W$ are the height and width of each slice, respectively. We adopt a pre-trained ResNet-50 network as the feature extractor, which applies a series of convolutional and pooling operations to extract spatial information from each slice. After global average pooling along the channel dimension, a unified-dimensional feature representation is generated. The resulting sMRI modality feature representation $F_{sMRI}$ is:

$$F_{sMRI} = ResNet50(X_{sMRI}) \in \mathbb{R}^{N \times d} \tag{5}$$

where $d$ denotes the dimension of the extracted features.

For the SNP modality, the input is a preprocessed and filtered SNP sequence, denoted as $X_{snp} \in \mathbb{R}^{N \times F}$, where $N$ is the number of SNP segments (instances), and $F$ is the feature dimension of each SNP. To model the dependencies among segments, we treat the input as a sequential signal and utilize a GRU network to extract sequence representations. The resulting SNP modality feature representation $F_{snp}$ is:

$$F_{snp} = GRU(X_{snp}) \in \mathbb{R}^{N \times d} \tag{6}$$

Finally, we have completed the initial feature extraction for both modalities, obtaining $F_{sMRI}$ and $F_{snp}$, respectively.

## 3.2. Multi-Instance Attention Module

We propose a novel **Multi-Instance Attention Module (MAM)**, which enables individualized and adaptive selection as well as objective scoring of SNP fragments, while also adapting the importance of sMRI slices.

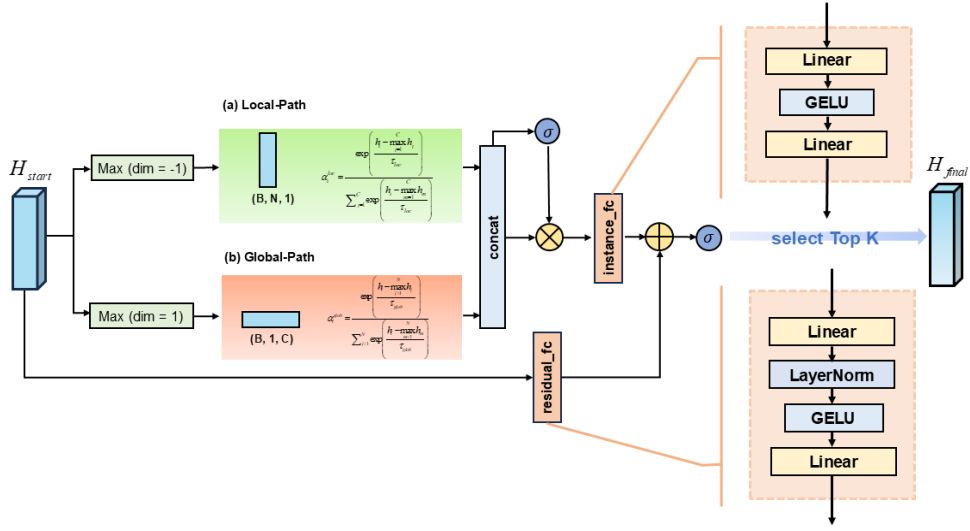

**Fig. 5**. Multi-Instance Attention Module (MAM）

As shown in Fig. 5, this module captures the heterogeneity across individuals in terms of genetic and pathological representations, and it identifies potential risk-associated SNP regions and key structural brain areas. MAM evaluates the important feature expressions of each individual under different modalities, significantly enhancing the model's multimodal fusion capability and interpretability. Its core workflow includes **dual-path attention computation**, **gated feature fusion**, and **Top-k instance selection**, with detailed implementation as follows:

In MAM, we input SNP features $F_{snp} = \{f_i^{snp} \mid i = 1, 2, \ldots, N\} \in \mathbb{R}^{N \times d}$ and sMRI features $F_{sMRI} = \{f_i^{sMRI} \mid i = 1, 2, \ldots, N\} \in \mathbb{R}^{N \times d}$ in parallel. To facilitate computation, the input features are first projected into a unified feature space of dimension $C$ through a linear layer, resulting in the mapped feature representations denoted as $H_{start} = \{h_i \mid i = 1, 2, \ldots, N\} \in \mathbb{R}^{N \times C}$. For $H_{start}$, MAM designs a **dual-path attention computation** to capture both local and global variations. Specifically, the **Local-Path** aggregates over the feature dimensions of each instance to capture salience within its feature space; the **Global-Path** aggregates along the instance sequence dimension to assess the overall importance of each fragment across the individual.

First, for each instance, the values across feature dimensions are normalized using softmax after subtracting the maximum value along the feature dimension. This yields the relative importance of each instance across all feature dimensions. The attention weight $\alpha_i^{loc}$ of the Local-Path is computed as:

$$\alpha_i^{loc} = \frac{\exp\left(\frac{h_i - \max_{j=1}^{C} h_j}{\tau_{loc}}\right)}{\sum_{j=1}^{C} \exp\left(\frac{h_i - \max_{m=1}^{C} h_m}{\tau_{loc}}\right)} \quad (7)$$

where $h_i$ is the feature vector of the current instance, $\max_{j=1}^{C} h_j$ and $\max_{m=1}^{C} h_m$ is the maximum value over its feature dimensions, and $\tau_{loc}$ is a learnable **temperature coefficient** to control the smoothness of the softmax.

Next, we obtain the representation $f_i^{loc}$ of each instance in the **Local-Path** by performing a weighted average:

$$f_i^{loc} = \gamma_{loc} \cdot \alpha_i^{loc} \cdot h_i \tag{8}$$

Where $\gamma_{loc} \in (0.3, 3.0)$ is an adaptive scaling factor to enhance the magnitude of the local features.

We then focus on capturing the global importance of the entire input instance sequence. Unlike the Local-Path, the **Global-Path** operates directly on the instances, preserving the semantic meaning of different features to obtain the overall representation. The attention weights $\alpha_i^{glob}$ in the Global-Path are calculated as follows:

$$\alpha_i^{glob} = \frac{\exp\left(\frac{h_i - \max_{j=1}^{N} h_j}{\tau_{glob}}\right)}{\sum_{j=1}^{N} \exp\left(\frac{h_i - \max_{m=1}^{N} h_m}{\tau_{glob}}\right)} \tag{9}$$

where $\max_{j=1}^{N} h_j$ and $\max_{m=1}^{N} h_m$ is the maximum value across the instance sequence, and $\tau_{glob}$ is a learnable global temperature coefficient to regulate the smoothness of global attention.

The aggregated coefficient $\alpha_i^{glob}$ reflects the relative importance of the *i-th* instance across the sequence in a specific modality. The representation $f_i^{glob}$ from the Global-Path is then computed by:

$$f_i^{glob} = \gamma_{glob} \cdot \alpha_i^{glob} \cdot h_i \tag{10}$$

where $\gamma_{glob} \in (0.3, 3.0)$ is an adaptive scaling factor to enhance the magnitude of the global features.

After computing the dual-path attentions, we obtain the local feature $\alpha_i^{loc}$ and global feature $\alpha_i^{glob}$. To effectively fuse both types of attention information, we introduce a **gating mechanism** to dynamically adjust their weight ratios. We concatenate the two types of features into a joint vector and calculate a gating coefficient (denoted as *gate*) for each instance using a gating network:

$$gate = \sigma\left(W_g \cdot concat\left(f_i^{loc}, f_i^{glob}\right)\right) \cdot \beta_g \tag{11}$$

where $W_g$ is the weight matrix of the gating network, $\sigma(\cdot)$ is the sigmoid activation function, and $\beta_g$ is a preset scaling factor (default set to 0.2) to enhance decision boundary clarity and gradient propagation efficiency.

The fused feature representation $\tilde{h}_i$ is computed by weighting the local and global features with the gating coefficient:

$$\tilde{h}_i = \left(gate \odot f_i^{loc} + (1 - gate) \odot f_i^{glob}\right) \cdot \beta_f \tag{12}$$

where $\beta_f$ (default set to 0.2) is used to further adjust the scale of the fused features.

After completing the gated feature fusion, we obtain the final fused representation $\tilde{h}_i$ containing both local and global information. To select the most representative instances, we adopt a **Top-k selection mechanism**. The goal of Top-k selection is to choose the most representative instances based on attention weights, thereby suppressing interference from redundant information. Each fused feature $\tilde{h}_i$ is weighted, and a dual-branch network computes the attention weights. The **main branch (instance_fc)** $W_a$ learns the importance of features, while the **residual branch (residual_fc)** $W_r$ provides stable supplementary signals. The outputs from both branches are passed through sigmoid activation and summed, followed by max-softmax normalization to obtain final attention weights $\hat{\alpha}_i$:

$$\hat{\alpha}_i = \beta_{com} \left( \frac{\sigma\left(\frac{W_a \tilde{h}_i}{\tau_{att}}\right) + \sigma\left(\frac{W_r h_i}{\tau_{att}}\right)}{\max_j \left[\sigma\left(\frac{W_a \tilde{h}_j}{\tau_{att}}\right) + \sigma\left(\frac{W_r h_j}{\tau_{att}}\right)\right]} \right) \quad (13)$$

Where $\tau_{att}$ is a learnable **attention temperature coefficient**, $\beta_{com}$ is a preset total **feature amplification factor** (default set to 10), and $\sigma(\cdot)$ denotes the **sigmoid** function.

Finally, based on these attention weights $\hat{\alpha}_i \in \mathbb{R}^{N \times 1}$, the Top-k instances are selected as the final feature set $H_{final} \in \mathbb{R}^{K \times C}$. A detailed discussion on the choice of $k$ is provided in the experiments.

$$H_{final} = \{h_i \mid i \in TopK(H_{start})\} \quad (14)$$

## 3.3. Dominant Modality Selection Module

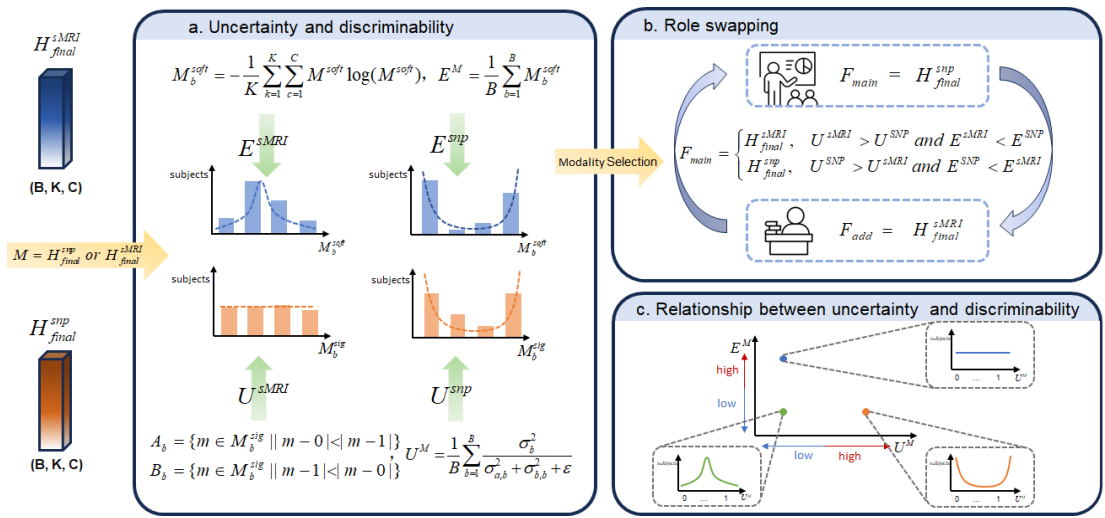

**Fig. 6**. Dominant Modality Selection Module. (a) Uncertainty and discriminability computation. (b) Role swap. (c) Relationship between uncertainty (entropy) and discriminability (U-shaped distribution). Although both Gaussian and U-shaped distributions have low entropy, the U-shaped

distribution exhibits higher discriminability due to its closeness to the ideal binary classification pattern.

Due to the significant differences in the diagnostic value of sMRI and SNP for AD, we designed the Dynamic Modality Selection Module (DMSM) to address this issue. This module effectively connects with the subsequent Contrastive Self-Distillation Module (CSM) and adaptively determines the roles of the dominant modality (teacher) and the auxiliary modality (student) in each training epoch, thus realizing a dynamic teacher-student role-swapping mechanism. The SNP features $H_{final}^{snp}$ selected by MAM and the sMRI features $H_{final}^{sMRI}$ are jointly fed into the DMSM.

As shown in Fig. 6, we model the reliability of features from the two modalities $M = H_{final}^{snp}$ or $H_{final}^{sMRI}$, $M \in \mathbb{R}^{K \times C}$. We assume that the more reliable modality should have the following characteristics: **(1) lower uncertainty in its posterior distribution and (2) higher discriminability of its posterior distribution.** The former is measured by the **entropy** of the posterior distribution, reflecting the modality's stability and confidence in classification results. The latter is quantified by a **scoring mechanism** designed to evaluate whether the feature distribution conforms to an ideal **U-shaped probability distribution** (similar to a $Beta(0.5, 0.5)$ distribution), indicating discriminability.

For discrete features, the entropy reaches its maximum when the probability distribution is uniform (i.e., each possible value has equal probability), indicating the highest level of uncertainty. In contrast, when the feature distribution follows a Gaussian or U-shaped distribution, the entropy is relatively lower, reflecting reduced uncertainty. Features with lower entropy typically correspond to higher classification confidence, which helps improve the stability of model performance. The specific calculation steps are as follows: First, the modality features of a single sample $M \in \mathbb{R}^{K \times C}$ are normalized using softmax along the feature dimension $C$, resulting in $M^{soft} \in \mathbb{R}^{K \times C}$. Then, the average information entropy $M_b^{soft}$ of the sample is calculated by averaging over all $K$ instances:

$$M_b^{soft} = -\frac{1}{K} \sum_{k=1}^{K} \sum_{c=1}^{C} M^{soft} \log(M^{soft}) \tag{15}$$

During training, the overall entropy (averaged over all $B$ samples) is taken to obtain the uncertainty of modality $E^M$:

$$E^M = \frac{1}{B} \sum_{b=1}^{B} M_b^{soft} \tag{16}$$

In a binary classification task (where 0 represents one class and 1 represents the other), the ideal scenario is that the modality features extracted by the model follow a U-shaped distribution, meaning most features are close to 0 or 1, with fewer around the middle. To quantify this property, we design a **U-shaped distribution scoring function** with the following steps. After applying a Sigmoid activation to the modality features $M \in \mathbb{R}^{K \times C}$, we obtain the transformed features $Q$. For each sample (i.e., the $b$-th data in the batch), its feature vector $Q_b$ is flattened into a one-dimensional vector, and two sets of elements are defined:

$$A_b = \{q \in Q_b \mid |q - 0| < |q - 1|\}$$
$$B_b = \{q \in Q_b \mid |q - 1| < |q - 0|\} \quad (17)$$

where $A_b$ represents all feature values closer to 0 in sample b, and $B_b$ represents those closer to 1.

The overall concentration score (averaged over all $B$ samples) is computed by:

$$U^M = \frac{1}{B} \sum_{b=1}^{B} \frac{\sigma_b^2}{\sigma_{a,b}^2 + \sigma_{b,b}^2 + \varepsilon} \quad (18)$$

Here, $U^M$ denotes the U-shape distribution score of the input modality features $M$, $\sigma_a^2$ is the total variance, $\sigma_{a,b}^2$ and $\sigma_{b,b}^2$ are variances of $A_b$ and $B_b$, respectively; $\varepsilon$ is a small positive constant to prevent division by zero. A higher score indicates the data distribution is more concentrated at the extremes, closer to a U-shaped distribution, which corresponds to stronger binary classification discriminability.

Based on these metrics, if a modality shows both lower uncertainty (lower entropy) and higher discriminability (higher U-shape score), it is assigned as the dominant modality $F_{main}$ (Teacher) for the current training epoch, while the other modality is assigned as the auxiliary modality $F_{add}$ (Student). If the opposite modality outperforms on both metrics, the roles are swapped accordingly:

$$F_{main} = \begin{cases} H_{final}^{sMRI}, & U^{sMRI} > U^{SNP} \text{ and } E^{sMRI} < E^{SNP} \\ H_{final}^{snp}, & U^{SNP} > U^{sMRI} \text{ and } E^{SNP} < E^{sMRI} \end{cases} \quad (19)$$

It is worth noting that when the two modalities exhibit conflicting indicators—such as one having higher uncertainty but also a higher U-shape score, or both metrics being lower for the same modality—this study defaults to choosing sMRI as the dominant modality, because sMRI directly reflects brain structural changes during AD progression, possessing higher physiological certainty and discriminability. This dynamic modality selection mechanism flexibly allocates teacher-student roles across different training epochs, effectively enhancing bidirectional co-updating between modalities.

After determining the teacher and auxiliary modality roles, we introduce a cross-modal attention mechanism to strengthen inter-modality interaction. Specifically, let the dominant modality feature be $F_{main} \in \mathbb{R}^{K \times C}$ and the auxiliary modality feature be $F_{add} \in \mathbb{R}^{K \times C}$. We take $F_{main}$ as the Query, and $F_{add}$ as the Key and Value, constructing an information flow across modalities through a multi-head self-attention module:

$$F_{cross} = MultiheadAttn(F_{main}, F_{add}, F_{add}) \quad (20)$$

where $F_{cross} \in \mathbb{R}^{K \times C}$ denotes the semantically enhanced representation of the auxiliary modality.

## 3.4. Contrastive Self-distillation Module

To further enhance cross-modal collaborative representation and reinforce feature alignment between modalities, we propose the **Contrastive Self-distillation Module (CSM)**, inspired by the teacher–student architecture in self-distillation. The core idea of CSM is to dynamically assign teacher and student roles, guiding the dominant modality $F_{main}$ (Teacher) to transfer high-confidence knowledge to the auxiliary modality $F_{add}$ (Student), while also allowing reverse corrections. CSM consists of three key components: **feature distillation, cross-modal contrastive learning,** and **classification loss.**

First, for the classification outputs of both modalities, the probability distribution from the dominant modality $F_{main}$ is used as a **soft label** to supervise the prediction of the auxiliary modality. The distillation loss is implemented using **Kullback–Leibler (KL) divergence**:

$$Z_t = classifier(LayerNorm(F_{main}))$$
$$Z_S = classifier(LayerNorm(F_{add})) \tag{21}$$

Here, $z_t$ and $z_s$ denote the global feature representations of the dominant and auxiliary modalities, respectively. The final KL-based distillation loss $L_{KL}$ is computed as the average over all samples in the batch:

$$L_{KL} = \frac{1}{B}\sum_{b=1}^{B} D_{KL}(soft\,max(z_t) \,||\, soft\,max(z_s)) \tag{22}$$

where $B$ is the batch size.

Next, we introduce an additional loss that jointly considers **modality uncertainty (entropy)** and **discriminability (U-shape distribution score)**. This term encourages the auxiliary modality to improve its representation quality and become more competitive against the dominant modality. The uncertainty-aware loss $L_{DMSM}$ is defined as:

$$L_{DMSM} = \lambda_a((E^{snp} + E^{sMRI}) - (U^{snp} + U^{sMRI})) \tag{23}$$

Where $E^{snp}$ and $E^{sMRI}$ denote the entropy of the SNP and sMRI modalities, respectively; $U^{snp}$ and $U^{sMRI}$ are their corresponding U-shape distribution scores, and $\lambda_a$ is a balancing weight (default set to 0.2).

Together, the above two components form the total **feature distillation loss**:

$$\mathcal{L}_{distill} = \lambda_b(\mathcal{L}_{KL} + \mathcal{L}_{DMSM}) \tag{24}$$

where $\lambda_b$ is the weight for distillation loss(default set to 0.1).

To minimize the feature discrepancy between modalities under the same semantic category, we introduce a **cross-modal contrastive loss** based on **cosine similarity**. This loss encourages feature representations of samples with the same label to be closer across modalities, while ensuring that representations of different-class samples remain distinguishable. For each training batch containing $B$ samples, we extract features from both the dominant and auxiliary modalities. The **cross-modal contrastive learning** is then defined as:

$$\mathcal{L}_{con} = 1 - \frac{1}{B}\sum_{b=1}^{B}\frac{F_{main} \bullet F_{add}}{\|F_{main}\|\|F_{add}\|} \tag{25}$$

Finally, the overall training objective of CSM integrates the **classification loss** $\mathcal{L}_{CE}$, **feature distillation loss** $\mathcal{L}_{distill}$, **and cross-modal contrastive learning** $\mathcal{L}_{con}$:

$$\mathcal{L} = \mathcal{L}_{CE} + \lambda_d \cdot \mathcal{L}_{distill} + \lambda_c \cdot \mathcal{L}_{con} \tag{26}$$

where $\mathcal{L}_{CE}$ is the standard cross-entropy classification loss, $\lambda_c$ is the weight for contrastive loss(default set to 0.1), and $\lambda_d$ is the weight for distillation loss(default set to 0.1).

## 4. Experiments

### 4.1. Datasets

The data used in this study were obtained from the ADNI database. This database is a large public database for AD (http://adni.loni.usc.edu/) and the data used include the ADNI-1, ADNI-GO, ADNI-2 and ADNI-3 cohorts. SNP data are based on Illumina 2.5M arrays and Illumina Omni Quad arrays. Image data are based on T1W1-3D-MP RAGE. detailed data are shown in Table.1.

**Table. 1**. Demographic Information

| Subgroup | Number | Gender(M/F) | Age | Mmscore |
|---|---|---|---|---|
| NC | 339 | 162/177 | 74.14±5.67 | 29.19±1.03 |
| AD | 124 | 68/56 | 76.50±7.80 | 23.02±2.16 |

SNP quality check and screening: we performed QC and screening on these genotype data using the PLINK software package (http://pngu.mgh.harvard.edu/purcell/plink/). QC consists of six parts: missing value processing, sex QC, MAF minimal allele test, HW Harwin equilibrium test, heterozygosity test and Screening. (1) For SNPs, SNP loci and samples with ≥2% deletions were removed. (2) Remove samples that are sexually inconsistent. (3 Minimum allele frequency (MAF) ≥5%. (4) Remove loci with p≤0.0001 in the Hardy-Weinberg equilibrium test (HWE) for SNPs. (5) Individuals with heterozygosity other than three times the standard deviation were removed. After QC this study did gemma mixed linear regression with age (age) and gender (gender) and the Brief Mental State Examination Scale (MMSE) as covariates (cov) to determine the association of each SNP with AD. (6) Using SNPs highly associated with AD and significant SNPs analyzed by gemma mixed linear regression in AlzData (AD database),

30 SNPs were screened as shown in Table 2 and used as the input data set for the study.

Table. 2. Top 30 Selected SNPs

| chr | rs ID | Position (bp) | Effect Allele | Other Allele | P value |
|---|---|---|---|---|---|
| 18 | rs2276269 | 45356000 | T | C | $2.70 \times 10^{-7}$ |
| 10 | rs3006968 | 36967697 | T | C | $4.07 \times 10^{-6}$ |
| 21 | rs2834098 | 33428176 | T | C | $8.06 \times 10^{-6}$ |
| 16 | rs7205641 | 52952111 | T | G | $1.11 \times 10^{-5}$ |
| 12 | rs1880845 | 104377662 | A | G | $1.60 \times 10^{-5}$ |
| 21 | rs2837284 | 40197271 | A | G | $1.91 \times 10^{-5}$ |
| 12 | rs11107229 | 76816892 | A | G | $2.56 \times 10^{-5}$ |
| 16 | rs2042416 | 13242472 | T | C | $2.58 \times 10^{-5}$ |
| 5 | rs1866374 | 71526718 | T | C | $3.02 \times 10^{-5}$ |
| 5 | rs11742315 | 55998350 | T | C | $4.67 \times 10^{-5}$ |
| 12 | rs12299627 | 104454300 | T | G | $6.13 \times 10^{-5}$ |
| 12 | rs12299724 | 104421138 | G | A | $7.87 \times 10^{-5}$ |
| 17 | rs227802 | 3289065 | A | G | $8.29 \times 10^{-5}$ |
| 16 | rs1994766 | 6954663 | C | T | $8.89 \times 10^{-5}$ |
| 15 | rs884483 | 68189363 | C | T | $8.92 \times 10^{-5}$ |
| 12 | rs15750 | 44864008 | T | C | $1.05 \times 10^{-4}$ |
| 16 | rs11077121 | 6976049 | A | G | $1.41 \times 10^{-4}$ |
| 12 | rs7975809 | 126254967 | G | A | $1.48 \times 10^{-4}$ |
| 7 | rs10487510 | 113089498 | C | T | $1.63 \times 10^{-4}$ |
| 4 | rs1364951 | 37695831 | C | A | $1.74 \times 10^{-4}$ |
| 7 | rs1476612 | 143999691 | T | C | $2.16 \times 10^{-4}$ |
| 21 | rs2822710 | 14806346 | T | G | $2.26 \times 10^{-4}$ |
| 10 | rs1412444 | 90992907 | A | G | $2.85 \times 10^{-4}$ |
| 11 | rs4910364 | 11472106 | A | G | $2.86 \times 10^{-4}$ |
| 5 | rs11958964 | 17826787 | A | G | $3.12 \times 10^{-4}$ |
| 2 | rs11682390 | 3166705 | G | A | $3.29 \times 10^{-4}$ |
| 15 | rs12148472 | 77018533 | C | T | AlzData |
| 16 | rs9934438 | 31012379 | A | G | AlzData |
| 19 | rs405509 | 50100676 | C | A | AlzData |
| 19 | rs439401 | 50106291 | T | C | AlzData |

sMRI data processing: We used the MRIcroGL tool to convert the image data from DCOM format to NII format, and then pre-processed the 3D brain images of all subjects with voxel-based texture analysis (VBM) using the SPM12 tool of MATLAB software, which mainly includes segmentation, alignment and spatial normalization. The steps are as follows. (1) Segment the sMRI data of the brain to get the 3D brain tissue maps such as gray matter. (2) 2D slicing of the 3D gray matter. (3) Due to the inconsistency in the size of the subject's images, it is necessary to remove part of the boundary that has no medical information. In this paper, the spatial resolution of all the data is adjusted to 256*256 slices before model training, and then center cropping is performed to finally obtain image slices of size 224*224. Finally, the slices in the range of 40-69 are used as the input data for the model.

## 4.2. Implementation details

The experiments in this paper were conducted on Ubuntu 22.04 based on the PyTorch 2.1.2 framework using an NVIDIA RTX 3090 GPU graphics card with 24GB of video memory. In this paper, the batchsize of MRI as well as SNP samples is set to 16. In this paper, the network is trained end-to-end using the Adam optimizer [30], with the initial learning rate set to 1e-4, weight decay to 1e-5, and epoch set to 200. Our GenDMR consists of an FG, a MAM, a DMSM and a CSM, and the FG consists of two feature extraction networks, GRU and ResNet-50, with the dimensions of the outputs of the two networks set to 1024. In this paper, four metrics are introduced for evaluating the model: Area Under Curve (AUC), Accuracy (ACC), precision (PRE), and specificity (SPE). To ensure fairness, in this paper, the parameter configurations of the comparative models are adjusted to the optimal form for training, and the performance of all models is evaluated using the average of five 50% discounted cross-validations.

## 4.3. Model comprisions

This paper compares three category of models. The first category is traditional machine learning models. The second category is the hybrid machine learning and deep learning models. The last category is advanced deep learning models. Specifically, the first class of models are XGBoostC + XGBoostR [31], RandomForesetC + RandomForesetR [32], and SVC + SVR [33]. The second class of models uses a mixture of machine learning and deep learning. Two state-of-the-art models for multimodal diagnosis were selected for comparison are VisionVision Transformer + XGBoostR [34] and DesNet + XGBoost [35]. The third category of models are deep learning models specialized for sMRI and SNP multimodal studies, which are IGIE [36] and IGNet [37], respectively. These models are all implemented based on their public code or described in their original papers and use the same training and testing datasets as in this study to ensure fairness in the comparison. The following is a brief description of the comparison models:

**XGBoost(C + R):** Two independent XGBoost models are used to model the features of sMRI and SNP, respectively. The sMRI modality is input into the classifier (XGBoostC) after time-dimensional maximum pooling and spreading, and the category probabilities are extracted as features. The SNP modality is also pooled and normalized and input into the regressor (XGBoostR), and the output regression scores are used as features. Finally, the two modal features are spliced in the channel dimension to complete the binary classification through a single-layer fully connected network.
**RandomForest(C + R):** Similar to XGBoost for sMRI and SNP are modeled using classifiers and regressors respectively. The category probabilities and regression scores are extracted as features, spliced and bicategorized by a fully connected layer. **SVC + SVR:** Similar to XGBoost, sMRI and SNP are modeled using SVM classifier and regressor respectively. The category probabilities and regression scores are extracted as features, which are spliced together to complete the binary classification via the fully connected layer. **VIT + XGBoostC:** The sMRI features were extracted using ViT and modeled by XGBoostC. SNP patterns were normalized and modeled by XGBoostR. The category probabilities and regression scores were output as feature representations, respectively, and the final splicing was done through a single fully connected layer to complete the classification. **DesNet+XGBoost:** The sMRI features were extracted using 3D DenseNet. SNP features are extracted using 1D CNN and combined with Layer Conductance for feature filtering to output a 100-dimensional genetic representation. The two modal features are spliced and downscaled by a linear layer, and the final classification is done by XGBoost. **IGIE:** The sMRI modality inputs 30 slices, and the layer-by-layer brain region features are extracted by masked residual convolution and flattened into a global representation. 30 pre-selected SNPs are input to the SNP modality, and gene dependencies are modeled using a position-less Transformer with an auxiliary classifier to enhance local features. The global features of the two modalities are fused with cross-modal attention and comparative learning, and the AD classification is finally completed by the full connectivity layer. **IGNet:** The sMRI spatial features were extracted using 3D convolution, and Transformer modeled SNP sequences. The two modal features are fused by element-by-element multiplication to finalize the AD classification.

As shown in Table.3, GenDMR achieved the overall optimal performance, as determined by paired t-tests ($p < 0.05$). Meanwhile, except for GenDMR, IGNet achieved superior results, followed by VIT + XGBoostR. For the former, this may be due to the learning of sMRI spatial information by IGNet. For the latter, this may be due to the fact that the attention mechanism of Transformer in VIT can better learn multimodal data features. Taken together, this suggests that GenDMR encodes and learns SNP spatial information, and it is reasonable to propose an attention-based MAM.

Table. 3. Model comprisions results

| Mean+std (%) | AUC | ACC | PRE | SPE |
|---|---|---|---|---|
| XGBoost (C + R) | 73.32±0.37 | 64.83±1.09 | 67.01±0.54 | 77.04±1.33 |
| RandomForest (C + R) | 80.14±0.08 | 64.84±0.01 | 65.65±0.00 | 81.39±0.42 |
| SVC + SVR | 87.03±2.55 | 80.01±0.88 | 88.74±0.28 | 96.85±0.47 |
| DesNet+XGBoost | 72.29±0.16 | 74.95±0.10 | 53.49±1.16 | **98.51±0.00** |
| VIT + XGBoostR | 87.45±0.27 | 82.54±0.76 | 75.95±1.05 | 93.88±0.08 |
| IGIE | 69.40±7.66 | 75.38±2.06 | 53.78±21.22 | 90.10±8.41 |
| IGNet | 91.92±0.83 | 87.09±0.32 | 91.59±0.34 | **98.51±0.00** |
| **Ours GenDMR** | **94.75±0.33** | **89.81±0.62** | **91.63±0.83** | 98.51±0.01 |

## 4.4 Ablation experiments

### 4.4.1. Module ablation comparison

Table. 4. Module ablation Experimental design

|  | MAM | DMSM | CSM |
|---|---|---|---|
| Case I | ✓ | × | × |
| Case II | × | ✓ | × |
| Case III | × | × | ✓ |
| Case IV | × | ✓ | ✓ |
| Full Model | ✓ | ✓ | ✓ |

We conduct additional experiments on three main innovation modules in the GenDSM model. These three main innovation modules are: the Multi-Instance Attention Module (MAM), the Dominant Modality Selection module (DMSM), and the Contrastive Self-Distillation Module (CSM). These experiments are designed to validate the independent role of each module as well as the synergistic effects between them.Table.4 shows the detailed design of the module ablation experiments.

As shown in Table.5, the Full Model achieves optimal performance on all metrics except PRE and SPE, which are slightly lower. Moreover, Case IV, which has a dynamic faculty-student interchange mechanism, is only lower than the Full Model in ACC and PRE. Therefore, we conclude that each module in the framework plays its own key role in optimizing the parameters for a limited number of samples, especially the Dynamic teacher-student role exchange mechanism, which is synergistic between CSM and

DMSM. This mechanism realizes adaptive migration of cross-modal knowledge by dynamically adjusting inter-modal teacher-student roles in conjunction with CSM. Ultimately, it breaks through the limitations of single modality or static integration.

Table. 5. Module ablation comparison results

| Mean+std (%) | AUC | ACC | PRE | SPE |
|---|---|---|---|---|
| Case I | 92.83±0.12 | 88.50±0.44 | 91.93±0.48 | **98.52±0.00** |
| Case II | 93.85±0.55 | 89.27±0.54 | 90.71±2.04 | **98.52±0.00** |
| Case III | 94.21±0.26 | 88.94±0.22 | 92.01±0.66 | 98.51±0.00 |
| Case IV | 94.22±0.05 | 89.26±0.11 | **92.39±0.58** | 98.51±0.00 |
| Full Model | **94.75±0.33** | **89.81±0.62** | 91.63±0.83 | 98.51±0.01 |

**4.4.2. SNP encoding method comparison**

Table. 6. SNP encoding method comparison results

| Mean+std (%) | AUC | ACC | PRE | SPE |
|---|---|---|---|---|
| One-hot | 93.52±2.48 | 88.61±2.3 | **92.21±1.94** | **98.52±0.07** |
| Ours | **94.75±0.33** | **89.81±0.62** | 91.63±0.83 | 98.51±0.01 |

The traditional One-hot coding approach for SNPs was compared with our approach. As shown in Table. 6, our coding approach achieved the highest in ACC, AUC. Specifically, AUC improved from 93.52% to 94.75% and ACC improved from 88.61% to 89.81%. In contrast, the values that One-hot coding lagged behind on AUC and ACC far exceeded its advantages on SPE and PRE. This suggests that One-hot coding, while better able to handle common negative samples, is weaker at recognizing complex positive classes. Therefore, the introduction of spatial information on chromosomes facilitates the deep learning model to recognize the differences of different SNPs.

**4.4.3. Modality ablation comparison**

Table. 7. Modality ablation comparison results

| Mean+std (%) | AUC | ACC |
|---|---|---|
| sMRI | 93.50±0.50 | 88.28±0.65 |
| SNP | 58.50±0.50 | 73.32±0.00 |
| **sMRI + SNP** | **94.75±0.30** | **89.81±0.62** |

In order to verify the effectiveness and stability of the overall architecture, we performed modal ablation experiments. The modal ablation experiments evaluated the classification performance of the three configurations, sMRI, SNP, and sMRI +SNP, respectively. As shown in Table.7, sMRI +SNP performs the best, closely followed by sMRI, while the first two are much better than SNP. This suggests that sMRI provides more critical information in AD classification. It is worth noting that sMRI +SNP performance is close to that of sMRI, mainly due to the fact that SNP contributes less independently to the classification, and its role is more focused on optimizing decision boundaries and reducing misclassification.

## 5. Sensitivity Analysis And Discussion

### 5.1. Sensitivity analysis

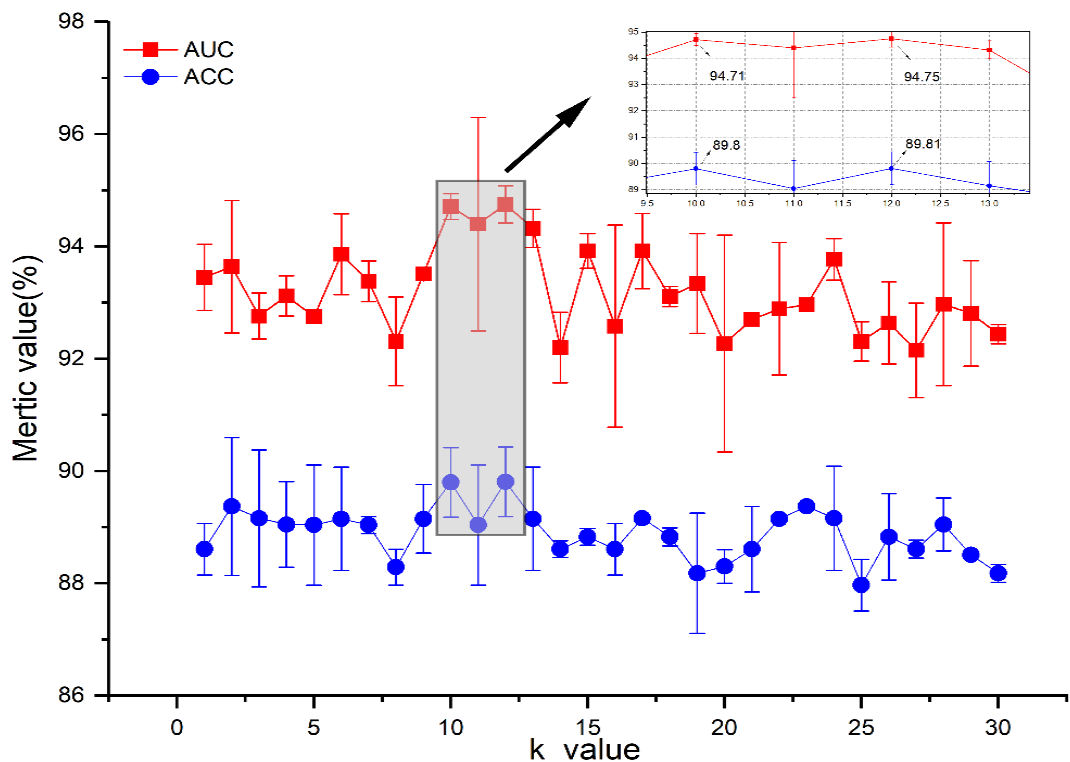

**Fig. 7**. The trend of average AUC and ACC with varying numbers of selected SNP features (k). The model achieves peak performance at k = 12, after which both metrics decline and become more unstable as k increases.

As shown in Fig 7, in order to assess the impact of the number of retained SNPs on the classification performance of the model, this study conducted a sensitivity analysis on the k value. Specifically, when the k value is small (1 ≤ k ≤ 12), the overall AUC and ACC of the model show an increasing trend with small fluctuations, which is in the performance improvement stage. And when k increases to 12, the AUC reaches 94.75% (±0.33%) and the ACC improves to 89.81% (±0.62%), both of which are the highest values in the whole interval. The increase in the number of features in the above intervals effectively improves the discriminative ability of the model, indicating that the addition of SNP features at this time has a high information gain.

When the value of k continues to increase to 13 and above, the model performance begins to fluctuate and gradually decreases, entering the performance saturation and decline phase. For example, the AUC slightly decreases to 94.32% (±0.34%) and the ACC is 89.15% (±0.92%) for k=13. Whereas, when k increased to 20, AUC decreased to 92.27% (±1.93%) and ACC to 88.30% (±0.30%). In particular, the AUC further decreased to 92.15% (±0.84%) and the ACC was 88.61% (±0.16%) for k=27. This phenomenon suggests that an excessive number of features leads to information redundancy and noise accumulation, which in turn affects the generalization performance of the model. After the value of k exceeds 20, the fluctuation of AUC and ACC of the model increases significantly, and the standard deviation increases, entering the stage of unstable performance. For example, the standard deviation of AUC at k=20 is as high as 1.93%, which is significantly higher than that at k=12 (0.33%). This indicates that with the introduction of redundant features, the model training process is interfered by noise, and the performance stability decreases.

Table .8. Coefficient of Variation Analysis for AUC and ACC (k = 1 to 30)

| Mean+std (%) | σ | μ | CV |
|---|---|---|---|
| AUC | 0.78% | 93.30% | 0.83% |
| ACC | 0.49% | 88.93% | 0.55% |

As shown in Figure Table 8, the coefficient of variation (CV=0.83%) of AUC from k=1 to k=30 is significantly higher than that of ACC (CV=0.55%), indicating that the change in the number of SNPs has a more significant effect on AUC. The multimodal fusion improves AUC by 36.25% (58.5% → 94.75%), while ACC improves it by only 16.51% (73.3% → 89.81%), which further corroborates that SNPs drive the performance enhancement through the decision boundary optimization (the core significance of AUC).

Notably, there is a significant difference in the impact of SNP features on AUC and ACC. When the number of SNP features varies from few to many, AUC shows an overall fluctuating trend of increasing and then decreasing, and peaks at k=12, while the fluctuation increases when there are too many SNP features showing the negative impact of redundant features. In contrast, ACC maintains a relatively smooth change. This difference suggests that SNP features have a key role in improving the model's classification discriminative ability (AUC), but have a limited effect on improving the overall accuracy (ACC). In summary, we choose k = 12 as the optimal number of features to ensure the best classification performance and the smallest fluctuation.

## 5.2. Discussion

### 5.2.1. Discussion on the Interpretability of SNPs

As emphasized by [38],[39], and [40], models with high classification accuracy are more likely to produce meaningful interpretations. Therefore, we considered the most important 12 of all SNPs to be potentially high-risk SNPs associated with AD. As shown in Fig 8, we use the SNP attention weight scores saved on the validation set for heat mapping. It can be seen that our model's evaluation of SNP importance is more focused. The 12 most important SNPs in descending order with their corresponding gene names are: VKORC1 (rs9934438), TLE3 (rs884483), NAV3 (rs11107229), TMEM132B (rs7975809), NUAK1 (rs12299724), RBL2 (rs7205641), PCP4 (rs2837284), TRAPPC12 (rs11682390), APOE (rs439401), IFNAR2 (rs2834098), GALNT18 (rs4910364), CKAP4 (rs12299627)。

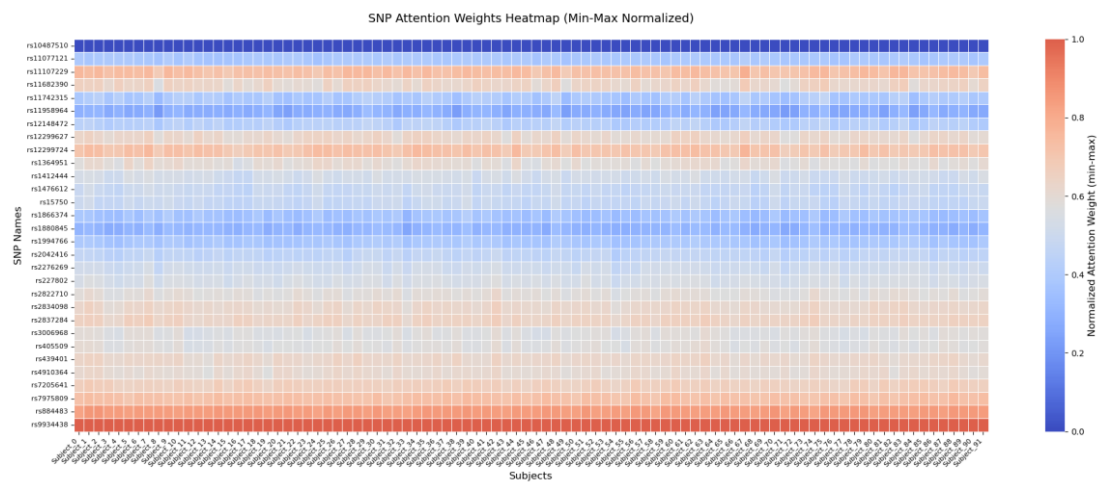

**Fig. 8**. Heatmap of attention weights for all SNPs. Each row represents a different SNP, and each column represents a different subject. The redder the color, the greater the attention the model gives to the corresponding SNP for a specific subject.

It should be emphasized that AD is characterized by amyloid (Aβ) deposition, hyperphosphorylation of Tau proteins and neuroinflammation, with which most of the genes localized in this study are closely related. The VKORC1 gene, which is of most interest in this model, is mainly involved in the function of vitamin K in the nervous system. It has been suggested that VKORC1 may influence the risk of AD development by regulating vitamin K metabolism [41]. In addition, a clinical study found that plasma levels of VKORC1 protein had good diagnostic properties for mild cognitive impairment (MCI) [42]. TLE3 has been shown to be associated with the development of memory T-cells [43] and to regulate lipid metabolism. And metabolic syndrome is one of the important risk factors for AD [44], so TLE3 may be indirectly involved in the development of AD through the metabolic pathway.NAV3 has been found to be closely associated with the pathological development of neurodegenerative diseases [45], and it is associated with elevated levels of peripheral blood inflammatory factors. This suggests that NAV3 may influence AD risk through inflammatory pathways [46].

It has been suggested that TMEM132B is associated with brain symptoms in AD patients [47], and several members of its gene family (TMEM132 family) have been shown to be highly associated with AD [48]. RBL2 is involved in neurodevelopmental processes [49], and it has been suggested that the RBL2 gene can be used as a biomarker for early diagnosis of AD [50]. NUAK1 expression is significantly up-regulated in brain tissues of AD patients and co-localized with neurofibrillary tangles (NFTs)[51], and it inhibits hyperphosphorylation of tau proteins, which is regarded as an important potential target for AD treatment [51]. PCP4, on the other hand, may contribute to AD pathology by regulating Aβ processing [52], and its expression level is significantly reduced in AD patients [53]. In an AD genome-wide association study, TRAPPC12 and its neighboring genes showed significant genetic associations with typical neuropathological features of AD, such as NFTs and cerebral amyloid angiopathy [54, 55].

The relationship between APOE and AD is considered to be one of the most central and well-studied in the field [56], the details of which will not be repeated here. What has been demonstrated is that increased expression of IFNAR2 is associated with microglial dysfunction [57]. Whereas microglial dysfunction is a central mechanism in AD etiology, many AD risk genes are highly and sometimes exclusively expressed by microglia [58]. In addition, GALNT18 is strongly associated with the rate of ventricular enlargement, which is an important hallmark of AD progression [59], and has been further suggested by several recent studies to have a possible role in AD pathogenesis [60, 61]. Finally, CKAP4, also known as P63. Although the involvement of CKAP4 in AD pathogenesis has not been reported, its potential association with AD has been confirmed by many studies [62]. For example, one study found that there were differences in the expression levels of CKAP4 in patients of different genders with AD [63].

### 5.2.2. Discussion on the Interpretability of sMRI

We identify the top 12 sMRI slices as the most critical regions highly associated with AD. As shown in Fig. 9, we average the attention scores across five validation runs to evaluate the relative importance of all slices. This provides a robust assessment of the model's focus on different sMRI slices in relation to AD diagnosis.

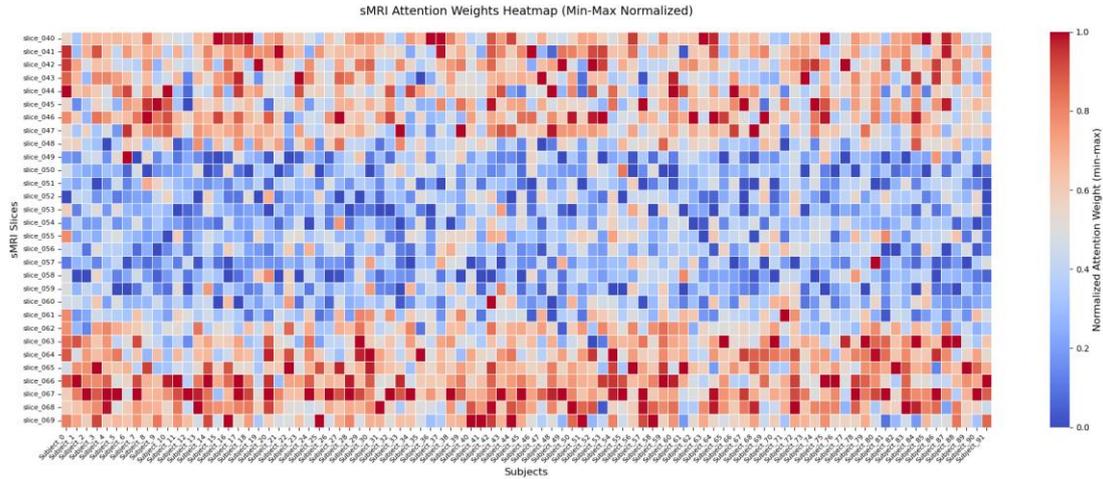

**Fig. 9**. Heatmap of attention weights for all sMRI slices. Each row represents a different sMRI brain slice, and each column represents a different subject. The redder the color, the greater the attention the model gives to the corresponding sMRI slice for a specific subject.

To further clarify the brain regions attended to by the model, we **identified the most discriminative brain region in each slice between the AD and NC groups based on gray matter maps** and the AAL90 brain atlas [64]. Specifically, the AAL90 atlas was first resampled to ensure alignment in resolution and spatial coordinates with each subject's gray matter map. Then, for each slice, we calculated the gray matter volume for each brain region and determined its proportion within the slice. The region with the highest proportion difference between the two groups was selected as the most discriminative region in that slice.

The identification process of the most discriminative brain region per slice is as follows:
1. For each slice, extract the 2D section of the AAL90-labeled gray matter map;
2. Within the non-background region, compute the total gray matter value for each labeled region;
3. Normalize the gray matter values to obtain the proportion of each region within the slice;
4. Calculate the difference in regional proportions between the AD and NC groups, and select the region with the largest difference as the most discriminative region for that slice. The region index is then mapped to its anatomical name using the atlas label file.

We use the most discriminative region in each slice as its structural representation, which provides biological plausibility for the model's attention mechanism and helps interpret the attended regions from a structural perspective. As shown in **Table 9**, the most discriminative regions across slices are predominantly located in the **Middle Frontal Gyrus Left (MFG.L)** [65], **Precuneus Left (PCUN.L)** [66], **Supramarginal

Gyrus Left (SMG.L) [67], and Middle Temporal Gyrus Left (MTG.L) [68], all of which have been previously implicated in Alzheimer's pathology.

Table.9. Max Difference Brain Regions Between AD and NC Focused by the Model

| sMRI_Slice | BrainRegion | Difference |
|---|---|---|
| slice_067 | MFG.L | 10.88 |
| slice_066 | PCUN.L | 8.42 |
| slice_068 | PCUN.L | 11.7 |
| slice_065 | SMG.L | 8.43 |
| slice_064 | SMG.L | 9.4 |
| slice_040 | MTG.L | 24.29 |
| slice_046 | MTG.L | 20.98 |
| slice_041 | MTG.L | 24.81 |
| slice_069 | PCUN.L | 11.54 |
| slice_042 | MTG.L | 31.61 |
| slice_047 | MTG.L | 21.02 |
| slice_063 | SMG.R | 10.04 |

## 6. Conclusion

This paper proposes GenDMR, a method that deeply encodes genetic features while distinguishing the roles of different modalities, aiming to diagnose AD and automatically identify potential risk genes. First, spatial information is incorporated into the SNP encoding process, and a multi-instance attention module is employed to adaptively quantify the risk contribution of each SNP. Second, to leverage the unique value of relatively weaker modalities, we design a **dynamic teacher-student role-swapping mechanism**, which consists of a **dominant modality selection module** and a **contrastive self-distillation module (CSM)**. This mechanism allows the model to adaptively select the most informative modality during training and enables bidirectional co-updating of modality roles.

Finally, we extensively investigate the impact of the hyperparameter *k* (the number of selected SNPs and sMRI slices) on model performance, analyze the unique contribution of SNPs compared to sMRI, and visualize attention scores across different SNPs. Our visualization demonstrates both **stability** and **focus**, leading to the identification of 12 potentially high-risk AD-related genes, including the well-known **APOE** and several emerging risk genes receiving increasing attention in recent studies. These results confirm that GenDMR possesses robust interpretability in the genetic analysis of AD and holds promise as a tool for assisting in the identification of disease-associated genes.